\documentclass[iop,apj,onecolumn,tighten]{emulateapj}
\usepackage{amsmath}

\begin{document}

\title{Three-Dimensional Moving-Mesh Simulations of Galactic Center Cloud G2}

\author{Peter Anninos}
\affil{Lawrence Livermore National Laboratory, Livermore, CA 94550, USA}
\author{P. Chris Fragile\footnotemark[1], Julia Wilson}
\affil{Department of Physics and Astronomy, College of Charleston, Charleston, SC 29424, USA }
\and
\author{Stephen D. Murray}
\affil{Lawrence Livermore National Laboratory, Livermore, CA 94550, USA}

\begin{abstract}
Using three-dimensional, moving-mesh simulations, we investigate the future evolution of the recently discovered gas cloud G2 traveling through the galactic center.  We consider the case of a spherical cloud initially in pressure equilibrium with the background.  Our suite of simulations explores the following parameters: the equation of state, radial profiles of the background gas, and start times for the evolution.  Our primary focus is on how the fate of this cloud will affect the future activity of Sgr A*.  From our simulations we expect an average feeding rate in the range of $5-19 \times 10^{-8} M_\odot~\mathrm{yr}^{-1}$ beginning in 2013 and lasting for at least 7 years (our simulations stop in year 2020).  The accretion varies by less than a factor of three on timescales $\le 1$ month, and shows no more than a factor of 10 difference between the maximum and minimum observed rates within any given model.  These rates are comparable to the current estimated accretion rate in the immediate vicinity of Sgr A*, although they represent only a small ($\lesssim 5$\%) increase over the current expected feeding rate at the effective inner boundary of our simulations ($r = 750 R_S \approx 10^{15}$~cm), where $R_S$ is the Schwarzschild radius of the black hole.  Therefore, the break up of cloud G2 may have only a minimal effect on the brightness and variability of Sgr A* over the next decade.  This is because current models of the galactic center predict that most of the gas will be caught up in outflows.  However, if the accreted G2 material can remain cold, it may not mix well with the hot, diffuse background gas, and instead accrete efficiently onto Sgr A*.  Further observations of G2 will give us an unprecedented opportunity to test this idea.  The break up of the cloud itself may also be observable.  By tracking the amount of cloud energy that is dissipated during our simulations, we are able to get a rough estimate of the luminosity associated with its tidal disruption; we find values of a few $10^{36} ~\mathrm{erg~s}^{-1}$.
\end{abstract}

\footnotetext[1]{KITP Visiting Scholar, Kavli Institute for Theoretical Physics, Santa Barbara, CA.}

\section{Introduction}

Sgr A*, the supermassive black hole at the center of the Milky Way galaxy, is the most underluminous black hole observed.  It has a bolometric luminosity of $\lesssim 10^{37}$ erg s$^{-1}$ \citep{narayan98}, which is less than $2 \times 10^{-8}$ of the Eddington luminosity, $L_\mathrm{Edd} = 5.2 \times 10^{44}$ erg s$^{-1}$, for a $M_{\rm BH} = 4.3 \times 10^{6} M_\odot$ black hole \citep{gillessen09}.  Part of the reason it is so underluminous is that it is starved for material.  At present, its accretion is fed by hot ($k_B T \approx 1.3$ keV) gas, possibly expelled in the winds of luminous young stars that are orbiting within its vicinity \citep{krabbe91,melia92,baganoff03}.  The electron number density of this gas, $n_e \approx 26$ cm$^{-3}$, suggests a Bondi feeding rate of $\approx 10^{-5} M_\odot~\mathrm{yr}^{-1}$ \citep{yuan03} at $r_\mathrm{Bondi} = GM_\mathrm{BH}/c_{s,b}^2 \approx 3.4 \times 10^{16}~\mathrm{cm}$.  However, measurements of polarization, along with Faraday rotation arguments, constrain the accretion rate at the black hole to be in the range $2\times10^{-9}<\dot{M} <2\times10^{-7} M_\odot~\mathrm{yr}^{-1}$ \citep{aitken00, bower03, marrone07}, suggesting it is difficult to actually get the gas down to the black hole.  This is consistent with a number of recent studies following the gas from these stellar winds, which have identified a number of physical processes that can significantly reduce the accretion efficiency, including expulsion due to hydrodynamic winds \citep{quataert04}, hang up due to residual angular momentum \citep{moscibrodzka06}, and heating due to electron thermal conduction \citep{shcherbakov10}.  Furthermore, some of the stars may produce winds that are too fast to be captured efficiently \citep{cuadra08}.

Within this environment, Sgr A* may occasionally enjoy a relative feeding frenzy.  Tidal disruption of stars passing close to the black hole is one way the accretion rate could be temporarily boosted.  Such an event may have been recently observed in a $z\approx0.35$ galaxy as the extremely luminous Swift transient event Sw 1644+57 \citep{bloom11,burrows11,zauderer11}.  Within the Milky Way, the star S2 has made the closest observed approach to Sgr A*, passing within 120 AU at its orbital pericenter \citep{gillessen09}.  Even so, to be disrupted, a star would need to pass within $\sim 1$ AU of Sgr A*.  Such close passes are only expected to occur approximately once every $10^4$ yr \citep[see][for a review]{alexander05}.

A second option, and the one that we explore in this paper, is tidal stripping of gas clouds passing through the galactic center.  \citet{gillessen12} recently discovered a gas cloud (dubbed G2) of $\sim3$ Earth masses on a highly eccentric orbit ($e = 0.9384$) approaching Sgr A* \citep{schartmann12}.  The cloud is expected to reach pericenter in 2013.5, at a distance of approximately 270 AU ($d_\mathrm{peri} = 4.0 \times 10^{15}~\mathrm{cm}$).  Because the self-gravity of the cloud is negligible \citep{burkert12}, it is certain to be tidally stretched.  The main questions are: 1) How much will hydrodynamic interactions with the background gas disrupt the cloud; 2) How much of the cloud material will be captured by Sgr A*; and 3) On what timescale will that captured material accrete?

To answer these questions we have performed three-dimensional hydrodynamic simulations of the cloud's infall.  Simulations of this cloud have been done previously \citep{burkert12,schartmann12}, but only in two dimensions.  Two-dimensional simulations offer the advantage that they allow one to cover the enormous dynamical range between the size of the cloud and the size of its orbit at reasonable computational expense.  However, such two-dimensional simulations are highly artificial.  One reason is that, in a Cartesian grid, a two-dimensional ``cloud'' has an infinite extent in its third dimension.  Furthermore, the enforced symmetry ignores the effects of tidal compression along this third dimension.  Finally, hydrodynamic instabilities, such as Kelvin-Helmholtz and Rayleigh-Taylor, are known to behave differently in two dimensions than in three \citep[e.g.][]{sharp84,fritts96,kane00,young01}.

We were able to get around the problem of the extreme dynamical range by using the moving mesh capability of {\em Cosmos++} \citep{anninos05,fragile12a}.  This allows our computational grid to follow the cloud along its orbit.  Because we only need to resolve the cloud and not the volume of its entire orbit, we are able to proceed in three dimensions at reasonable computational expense.

The paper is organized as follows: In Section \ref{sec:cloud}, we review the most pertinent details about the cloud's structure, internal properties, orbital parameters, and environment.  These are necessary for setting up the initial numerical problem, and also for estimating the relevant timescales for this problem, which we do in Section \ref{sec:timescales}.  This provides an opportunity to discuss the physical processes that will play a role in the disruption of the cloud as it passes pericenter.  As it turns out, many of the timescales for these processes are shorter than or comparable to the orbital time.  The competition among physical processes illustrates why a numerical approach is essential.  In Section \ref{sec:results}, we present the results of our three-dimensional, moving-mesh simulations following the cloud G2 through pericenter passage.  We summarize our findings in Section \ref{sec:conclusions}.  Additionally, since this is our first introduction of moving meshes in {\em  Cosmos++}, we provide a brief description of the equations in Appendix \ref{sec:mesh}.

\section{Cloud Properties}
\label{sec:cloud}

\citet{gillessen12} detected the dense, dusty, and ionized gas clump, G2, with a dust temperature of 550 K and a gas temperature of order $10^4$ K, embedded in the diffuse $10^8$ K gas of the galactic center. They adopt an effective, spherical cloud radius in 2011.3 of 15 mas, which corresponds to $R_c = 1.875\times10^{15}$ cm.  The observed Br$\gamma$ luminosity implies a cloud density of $\rho_c \approx 6.1\times10^{-19}f_V^{-1/2}$ g cm$^{-3}$, with a corresponding cloud mass of $M_c \approx 1.7\times10^{28}f_V$ g, or approximately 3 Earth masses, where $f_V \le 1$ is the volume filling factor. We will use this simple spherical approximation along with a filling factor $f_V = 1$ and a uniform cloud density to initialize our simulations, even though observations suggest that G2 might, in fact, be the compact head of a larger, more diffuse distribution of cold gas \citep{burkert12}.  Since the cooling timescale of the cloud is much shorter than its orbital period (see Section \ref{sec:timescales}), cooling plus ionization by the strong UV field of the central cluster of massive stars should combine to maintain the cloud temperature at $T_c = 10^4$ K, which we take as its initial value, although we also explore two non-isothermal models that are allowed to evolve away from this temperature.  Along with the ideal gas law, these parameters completely specify the cloud properties.

We consider two models for the background medium.  The first is based on the background parameters used in \citet{burkert12} and \citet{schartmann12}, i.e.
\begin{equation}
\rho_b = 1.3 \times 10^{-21} \eta \left( \frac{d}{d_{1995}} \right)^{-1} ~\mathrm{g~cm}^{-3}
\label{eq:rho_back_1}
\end{equation}
\begin{equation}
T_b = 1.2 \times 10^8 \left( \frac{d}{d_{1995}} \right)^{-1} ~\mathrm{K}.
\label{eq:T_back_1}
\end{equation}
where $d_{1995} = 10^4 R_S$ was the approximate distance of the cloud from Sgr A* in 1995.5 and $R_S = 2 G M_{\rm BH}/c^2 = 1.3 \times 10^{12}$ cm is the Schwarzschild radius of the black hole.  We have assumed a mean molecular weight of $\mu = 0.614$.  The parameter $\eta \le 1$ is meant to account for any unresolved X-ray sources contributing to the observed background X-ray luminosity \citep{sazonov12}.  In practice, we use this coefficient to adjust the background density to ensure that the background is initially in pressure equilibrium with the cloud.

One problem with this background is that it is convectively unstable, as the entropy, $S = T/\rho^{2/3}$, decreases radially outward.  In order to prevent the cloud from being disrupted by this convectively unstable atmosphere, we follow the same procedure that \citet{burkert12} and \citet{schartmann12} used to artificially stabilize it.  This starts by additionally evolving a passive tracer field $\cal{T}$.  The tracer is initialized to the same value as the cloud density inside the cloud and zero in the background.  This allows us to later distinguish between zones with pure atmospheric gas (${\cal T}/\rho < 10^{-4}$) from those with significant cloud material (${\cal T}/\rho \ge 10^{-4}$). Also, because the units we adopt in our calculations normalize the total cloud mass to unity, the value of the tracer field is a measure of the zonal mass relative to the total initial cloud mass.  At the end of each computational cycle, zones identified as pure atmosphere have their densities and temperatures reset to the background values given in equations (\ref{eq:rho_back_1}) and (\ref{eq:T_back_1}) and their velocities reset to zero.  We tested other values ($10^{-3}$ and $10^{-5}$) for the tracer floor and found that the mass accreted onto the black hole increases only by about 3\% with each factor of 10 that the tracer floor is lowered.

The second background we consider is based on a more careful fit to the galactic center environment profile  from \citet{yuan03}, designed to match {\em Chandra} X-ray observations.  This gives a background of the form
\begin{equation}
\rho_b = 1.3 \times 10^{-21} \eta \left( \frac{d}{d_{1995}} \right)^{-1.125} ~\mathrm{g~cm}^{-3}
\label{eq:rho_back_2}
\end{equation}
\begin{equation}
T_b = 10^8 \left( \frac{d}{d_{1995}} \right)^{-0.75} ~\mathrm{K}.
\label{eq:T_back_2}
\end{equation}
Unlike the atmosphere in equations (\ref{eq:rho_back_1}) and (\ref{eq:T_back_1}), this atmosphere is convectively stable.  Otherwise this background is implemented in the same way as the first one.

In both backgrounds we assume hydrostatic equilibrium within the gravitational potential of Sgr A*. This is certainly an oversimplification. Many physical processes are likely to play a role in determining the detailed structure of the galactic center environment.  However, as there is no detailed data available at this time to more faithfully model this, we focus only on the dynamical interaction between the cloud and these simple atmospheres.

We assume that the cloud is following a Keplerian orbit.  We start the cloud either from 1995.5, shortly before its first detection, or 1944.6, the date corresponding to the apocenter of its current orbit \citep{burkert12}.  In both cases the cloud is assumed to initially be in pressure equilibrium with the background.

\section{Timescales}
\label{sec:timescales}

In this section we consider the timescales associated with various physical processes that may act within the cloud or between the cloud and the background.  We do this to estimate their relative importance to the evolution of G2.  All estimates are scaled to the parameter values we use at the start of most of our simulations.  This corresponds to the cloud position in 1995.5, although we use the size and mass of G2 as measured in 2011.3.  This is appropriate since we initialize all of our simulations using the 2011.3 cloud properties.  Thus, the timescales we calculate give us a good estimate of the relative importance of each of these processes near the start of our typical simulations.

First we consider radiative cooling.  The associated timescale is  
\begin{eqnarray}
\tau_\mathrm{cool} & = & \frac{1.5 n_c k_B T_c}{\Lambda n_c^2} \nonumber \\
& \approx & 2.3 \times 10^4 f_V^{1/2} \left(\frac{\mu}{0.614}\right) \left(\frac{T_c}{10^4~\mathrm{K}} \right) \left(\frac{\Lambda}{\Lambda^*} \right)^{-1} \left( \frac{\rho_c}{\rho_c^*}\right)^{-1}  ~\mathrm{s},
\end{eqnarray}
where the numerator is the energy density of the cloud (assuming a gamma-law equation of state with adiabatic index $\Gamma = 5/3$), the denominator is the volume emissivity, $n_c = \rho_c/(\mu m_H)$ is the number density of the cloud, $m_H$ is the mass of hydrogen, $k_B$ is Boltzmann's constant, and $\rho_c^* = 6.1 \times 10^{-19}~\mathrm{g~cm}^{-3}$ is the initial cloud density used in our simulations.  To get a general estimate of the cooling timescale, we use the same cooling rate $\Lambda^* = 3 \times 10^{-22}~\mathrm{erg~cm^3~s^{-1}}$ as in \citet{burkert12}.  Since our estimated cooling timescale is much shorter than the orbital period of $\tau_\mathrm{orb} = 138$ yr, we are justified in assuming that the cloud will maintain its initial temperature of $T_c = 10^4$ K.

Small gas clouds like G2, embedded in a hot environment like the galactic center, will lose mass due to evaporation as a result of thermal conduction.  Whenever radiation and magnetic field effects are unimportant, the evaporation timescale, in the case when the heat flux reaches its limiting value (saturated state), can be estimated from \citep{cowie77}
\begin{eqnarray}
\tau_\mathrm{evap} & = & \frac{M_c}{\dot{M}_\mathrm{evap}} \nonumber \\
 & \approx & 73 \left(\frac{\rho_c}{\rho_c^*} \right) \left( \frac{\rho_b}{\rho_b^*}\right)^{-1} \left(\frac{R_c}{R_c^*} \right) \left(\frac{T_b}{1.2 \times 10^8~\mathrm{K}} \right)^{-1/2} ~\mathrm{yr}.
\end{eqnarray}
where 
\begin{equation}
\dot{M}_\mathrm{evap} \approx 1.7 \times 10^{19} \left( \frac{\rho_b}{\rho_b^*}\right) \left(\frac{R_c}{R_c^*} \right)^2 \left(\frac{T_b}{1.2 \times 10^8~\mathrm{K}} \right)^{1/2} ~\mathrm{g~s}^{-1}
\end{equation}
is the saturated mass loss rate due to evaporation,  $\rho_b^* = 1.3 \times 10^{-21} ~\mathrm{g~cm^{-3}}$ is the background density at $d_{1995} = 10^4 R_S$ for our default background, and $R_c^* = 1.875 \times 10^{15}~\mathrm{cm}$ is the initial radius of our simulated cloud.
This evaporation timescale is somewhat shorter than the orbital period. It is, therefore, unclear if G2 could traverse the full distance from its apocenter.  Further, it seems clear that once G2 breaks up as it approaches pericenter, the resulting fragments will likely dissolve quickly.

G2 will also experience ram pressure as it traverses the galactic center.  Ram pressure will compress the cloud on its leading edge and slow its motion relative to the background gas \citep{murray04}. The importance of ram pressure, compared to the thermal pressure of the surrounding gas, is given by the ratio $v_\mathrm{rel}^2/c_{s,b}^2$, where 
\begin{eqnarray}
c_{s,b} & = & \left( \frac{k_B T_b}{\mu m_H} \right)^{1/2} \nonumber \\
 & = & 1.3 \times 10^8 \left(\frac{\mu}{0.614}\right)^{-1/2} \left( \frac{T_b}{1.2 \times 10^8~\mathrm{K}} \right)^{1/2} ~\mathrm{cm~s^{-1}}
\end{eqnarray}
is the sound speed of the background gas and 
\begin{eqnarray}
v_\mathrm{rel} & \approx & \left[ 2 G M_\mathrm{BH} \left(\frac{1}{d_f} - \frac{1}{d_i}\right) \right]^{1/2} \nonumber \\
 & = & 2.8 \times 10^8 \left(\frac{M_\mathrm{BH}}{4.3 \times 10^6 M_\odot} \right)^{1/2} \left[ \left(\frac{d_f}{d_{1995}}\right)^{-1} - \left(\frac{d_\mathrm{apo}}{d_{1995}} \right)^{-1} \right]^{1/2}~\mathrm{cm~s^{-1}}
\end{eqnarray}
is the relative velocity between the cloud and the background.  For simplicity, we utilize the free-fall formula.  Noting that our background is stationary, we have $v_\mathrm{rel} = v_{c}$.  Close to apocenter ($d_\mathrm{apo} =  1.3 \times 10^{17}~\mathrm{cm}$), where $v_\mathrm{rel}=v_c \approx 0$, we see that gas pressure dominates and ram pressure is unimportant, while at the cloud's 1995.5 location, ram pressure begins to dominate.

Since the self-gravity of G2 is not important, the cloud will also be subject to tidal forces from the background potential.  Tidal forces will stretch the cloud in the direction of the black hole and squeeze it perpendicular to this direction. The net result is to redistribute the cloud along its orbital path. The acceleration of this stretching is given by
\begin{equation}
a_\mathrm{ts} = \frac{2 R_c G M_\mathrm{BH}}{d^3}~.
\end{equation}
This gives a timescale for tidal stretching of 
\begin{eqnarray}
\tau_\mathrm{ts} & \approx & \left( \frac{2 R_c}{a_\mathrm{ts}} \right)^{1/2} \nonumber \\
 & = & 2.0 \left(\frac{M_\mathrm{BH}}{4.3 \times 10^6 M_\odot} \right)^{-1/2} \left( \frac{d}{d_{1995}} \right)^{3/2}~\mathrm{yr}.
\end{eqnarray}
From this we expect the cloud to quickly be significantly tidally distorted.

At the surface of the cloud, we have two different density fluids moving relative to one another.  This makes the surface susceptible to the Kelvin-Helmholtz instability.  This instability will disrupt the surface of the cloud on a timescale
\begin{eqnarray}
\tau_\mathrm{KH} & = & \left(k v_\mathrm{rel}\right)^{-1} \left(\frac{\rho_\mathrm{c}}{\rho_\mathrm{b}}\right)^{1/2} \nonumber \\
 & \approx & 9.0 \left(\frac{R_c}{R_c^*}\right) \left(\frac{v_\mathrm{rel}}{2.8 \times 10^8 ~\mathrm{cm~s^{-1}}}\right)^{-1} \left(\frac{\rho_c}{\rho_c^*}\right)^{1/2} \left(\frac{\rho_b}{\rho_b^*}\right)^{-1/2} ~\mathrm{yr},
\label{eqn:kh}
\end{eqnarray}
where $k$ is the wavenumber of the perturbation.   We assume $k \approx 1/R_c$ to be the most disruptive wavenumber.  This estimate suggests Kelvin-Helmholtz disruption may become important, especially as the cloud approaches pericenter.

As the high-density cloud is ram-pressure decelerated by the low-density background environment, its leading edge will be subject to disruption by the Rayleigh-Taylor instability.  In the linear regime, this instability grows on a timescale of \citep{murray93}
\begin{eqnarray}
\tau_\mathrm{RT} & \approx & \left[\frac{M_{c}}{\pi R_{c}^{2} k \rho_{b}v_{rel}^2}\left(\frac{\rho_{b}+\rho_{c}}{\rho_{c}-\rho_{b}}\right) \right]^{1/2} \nonumber \\
 & \approx & 11 \left( \frac{M_c}{1.7 \times 10^{28} ~\mathrm{g}}\right)^{1/2} \left(\frac{R_c}{R_c^*}\right)^{-1/2} \left(\frac{\rho_b}{\rho_b^*}\right)^{-1/2} \left(\frac{v_\mathrm{rel}}{2.8 \times 10^8 ~\mathrm{cm~s^{-1}}}\right)^{-1} ~\mathrm{yr}.
\label{eqn:rt}
\end{eqnarray}
assuming the limit $\rho_c \gg \rho_b$.  Again the most destructive wavelength is $k \approx 1/R_c$.  This timescale is comparable to the Kelvin-Helmholtz timescale indicating both processes may act on the cloud simultaneously.

In summary, we find
\begin{equation}
\tau_\mathrm{cool} \ll \tau_\mathrm{ts} \lesssim \tau_\mathrm{KH} \lesssim \tau_\mathrm{RT} \ll \tau_\mathrm{evap} \lesssim \tau_\mathrm{orb} ~.
\end{equation}
The wide array of dynamical processes and timescales illustrate why realistic numerical simulations are so crucial.  The only process that may be important that we do not capture in this work is thermal conduction, which in any case is not expected to be a factor until late times.

\section{Simulations}
\label{sec:results}

The simulations use a Cartesian grid with a starting size of $14 R_c^*$ on each side, resolved with $256$ zones in each dimension, for a total of $256^3$ zones.  Doing so, we achieve an initial linear resolution of $\Delta x,y,z = 1.0 \times 10^{14} ~\mathrm{cm}$, comparable to that used in the 2D simulations of \citet{schartmann12}.  The fact that we can achieve a resolution in 3D comparable to that of a 2D simulation while covering the same effective volume illustrates the tremendous advantage of using a moving mesh.  

For a relatively simple case like the orbital motion of a gas cloud, it is most convenient to fix the motion of the mesh using the Keplerian velocity, rather than try to have the mesh dynamically adjust to the cloud's motion.  Thus, we set the grid velocity (as defined in Appendix \ref{sec:mesh}) to $V_g^{x,y}(t) = \zeta v_{c,\mathrm{Kep}}^{x,y}(t)$, where $v_{c,\mathrm{Kep}}$ is the Keplerian velocity vector of the cloud (assumed to lie in the $x$-$y$ plane) and $\zeta \lesssim 1$ is a parameter that allows us to account for the moderate slowing of the cloud due to ram-pressure deceleration.  We find $\zeta = 0.98$ works well here.  In the $z$ direction, the mesh is given a small velocity toward the $z=0$ plane with a magnitude $\vert V_g^z \vert = 5 R_c^*/(t_\mathrm{peri} - t_\mathrm{start})$, where $t_\mathrm{peri} = 2013.5$ and $t_\mathrm{start} = 1995.5$ or $1944.6$.  This compensates for the vertical slimming of the cloud due to tidal stretching and allows us to maintain an approximately constant resolution across the cloud in the $z$ direction.  This vertical grid compression stops at $t_\mathrm{peri}$.  As the cloud passes pericenter, the grid motion in the $x$-$y$ plane also changes, as we now wish to keep both the cloud and the black hole on the grid for the remaining evolution.  To accomplish this, the $x$ and $y$ components of the grid velocity change to the following forms at $t = 2014.7$ and $t = 2014$, respectively:
\begin{eqnarray}
V_g^x & = & v_{c,\mathrm{Kep}}^x(t) \frac{x_\mathrm{max}-x}{x_\mathrm{max}-x_\mathrm{min}}\\
V_g^y & = & v_{c,\mathrm{Kep}}^y(t) \frac{y_\mathrm{max}-y}{y_\mathrm{max}-y_\mathrm{min}}
\end{eqnarray}
where $x_\mathrm{max}$ and $y_\mathrm{max}$ are the maximum $x$ and $y$ grid positions during the time this grid velocity is implemented.  This velocity field holds the ``upper right'' edge of the grid fixed, retaining the black hole on the grid, while allowing the ``lower left'' edge to stretch to follow the cloud.  During this phase, the linear resolution of the grid decreases.  We terminate the simulations at $t_\mathrm{stop} = 2020$, with a linear resolution of $\Delta x = 2.2 \times 10^{14} ~\mathrm{cm}$ and $\Delta y = 1.5 \times 10^{14} ~\mathrm{cm}$.  The final $z$ resolution is $\Delta z = 6.3 \times 10^{13} ~\mathrm{cm}$.

The boundary conditions on the edges of the grid are set to outflow conditions (all ghost zone quantities are set equal to the values of their nearest internal-zone neighbor, except that the velocity component normal to the boundary is set to zero if it points onto the grid).  Similar to \citet{burkert12} and \citet{schartmann12}, we also introduce a spherical ``accretion radius'' centered on the black hole, which is located at the origin of our coordinate system.  All gas flowing into this region is treated as being accreted.  Each cycle, the amount of tracer material entering this region is tallied, and then all hydrodynamic variables inside it are reset to the static atmosphere solution.  For all of our reported simulations, the accretion radius is set to $r_\mathrm{acc} = 750 R_S$, half the value used in \citet{schartmann12}, although we saw only modest differences (19\%) in our measured mass accretion rate when we used the larger value.  

For this problem, we ignore the effects of the cloud and the background gas on the gravitational potential, and instead simply consider the (fixed) potential of the black hole.  For this work we use the pseudo-Newtonian, \citet{paczysky80} potential
\begin{equation}
\phi = \frac{GM_\mathrm{BH}}{r - R_S}~,
\end{equation}
although at the distances we are considering, there is no practical difference between this and a standard Newtonian potential for a point mass.

We experiment with three different equations of state: isothermal, isentropic, and polytropic. The first two are special cases of the polytropic equation of state, which specifies the pressure in the form $P=\kappa\rho^\Gamma$, where $\kappa$ is the polytropic constant, $\rho$ is the gas density, and $\Gamma$ is the adiabatic index.  Our default EOS is the isothermal option with $\Gamma = 1$.  This EOS, together with the perfect gas law ($P = k_B \rho T/\mu m_H$), ensures the cloud remains at a constant temperature. This can be thought of as representing a perfect balance between the heating of the cloud by the UV background and the radiative cooling of the cloud itself.  For the isentropic case, we set $\Gamma = 5/3$, which allows the temperature to vary, but $\kappa$ is held fixed. 

We also consider a third, more generic, polytropic gas with $\Gamma = 5/3$, but with $\kappa$ allowed to vary.  This introduces an additional degree of freedom to the gas equations in the form of internal energy [equation (\ref{eqn:av_en_covC})]. The manner in which the pressure is recovered also differs in this case from the previous two options.  For an isentropic (and isothermal) gas, the pressure is recovered directly from $P= \kappa\rho^\Gamma$, and the internal energy of the gas is not needed.  By contrast, the polytropic option derives its pressure from $P = (\Gamma-1)e$, where $e$ is the independently-evolved internal energy.  For purely adiabatic flows, polytropic and isentropic equations of state with the same $\kappa$ and $\Gamma$ should give nearly identical results, as the pressure will change in time according to the adiabatic scaling.  If there are shocks in the system, however, entropy is not conserved and the two equations of state may produce significantly different solutions. 

The equations of state are applied only to the cloud.  The thermodynamic state of the background gas is specified by one of the two hydrostatic solutions in Section \ref{sec:cloud}, together with the ideal gas law for the pressure. So long as cloud and background gases do not mix, one can apply the two procedures unambiguously. However, in reality cloud and background materials do mix as the cloud moves through the mesh, so a prescription for calculating a reasonable thermodynamic state for regions containing both gas and cloud materials must be provided. For our calculations we have chosen to impose a relaxation scheme that sets the temperature of mixed zones to a mass weighted combination of background and cloud temperatures. In practice this comes to: $T_\mathrm{mix} = ({\cal{T}}/\rho) T_c + (1-{\cal{T}}/\rho) T_g$, where $T_c$ is the cloud temperature calculated from the appropriate equation of state, $T_g$ is the background temperature calculated based on the zone position relative to galactic center, and ${\cal{T}}/\rho$ is the mass fraction of cloud material in the zone. We define mixed zones as those regions where $10^{-4} \le {\cal T}/\rho \le 0.8$. Once the temperature is calculated in this fashion, the pressure follows from the ideal gas law.

Table \ref{tab:models} summarizes the five models we consider in this paper, exploring three main parameters: equation of state, background model, and start date.  In Section \ref{sec:general}, we describe our generic results and compare with the 2D simulations of \citet{schartmann12}.  In Section \ref{sec:eos} we compare results from the different equations of state.  In Section \ref{sec:background}, we explore our two different backgrounds.  In Section \ref{sec:start}, we consider the two different start dates.  Finally in Sections \ref{sec:accretion} and \ref{sec:luminosity}, we study the mass accretion and luminosity histories of our simulations.

\begin{deluxetable}{llcclc}
\tablecaption{Model Parameters \label{tab:models}}
\tablewidth{0pt}
\tablehead{
\colhead{Model} & \colhead{EOS} & \colhead{$\Gamma$} & \colhead{Bkgd} &
\colhead{$t_\mathrm{start}$} & \colhead{$\langle \dot{M} \rangle$\tablenotemark{a}} \\
 & & & & & ($10^{-8}~M_\odot~\mathrm{yr^{-1}}$)
}
\startdata
cc\_i1\_b1\_95  & isothermal & 1   & 1 & 1995.5 & $5.18 \pm 2.25$ \\
cc\_i53\_b1\_95 & isentropic & 5/3 & 1 & 1995.5 & $7.75 \pm 6.15$ \\
cc\_p53\_b1\_95 & polytropic & 5/3 & 1 & 1995.5 & $5.37 \pm 4.08$ \\
cc\_i1\_b2\_95  & isothermal & 1   & 2 & 1995.5 & $4.64 \pm 2.19$ \\
cc\_i1\_b1\_44  & isothermal & 1   & 1 & 1944.6 & $19.0 \pm 12.1$
\enddata
\tablenotetext{a}{Mass accretion rate, time averaged from when accretion started until $t_\mathrm{stop} = 2020$.}
\end{deluxetable}

\subsection{General Results}
\label{sec:general}

\begin{figure}
\plotone{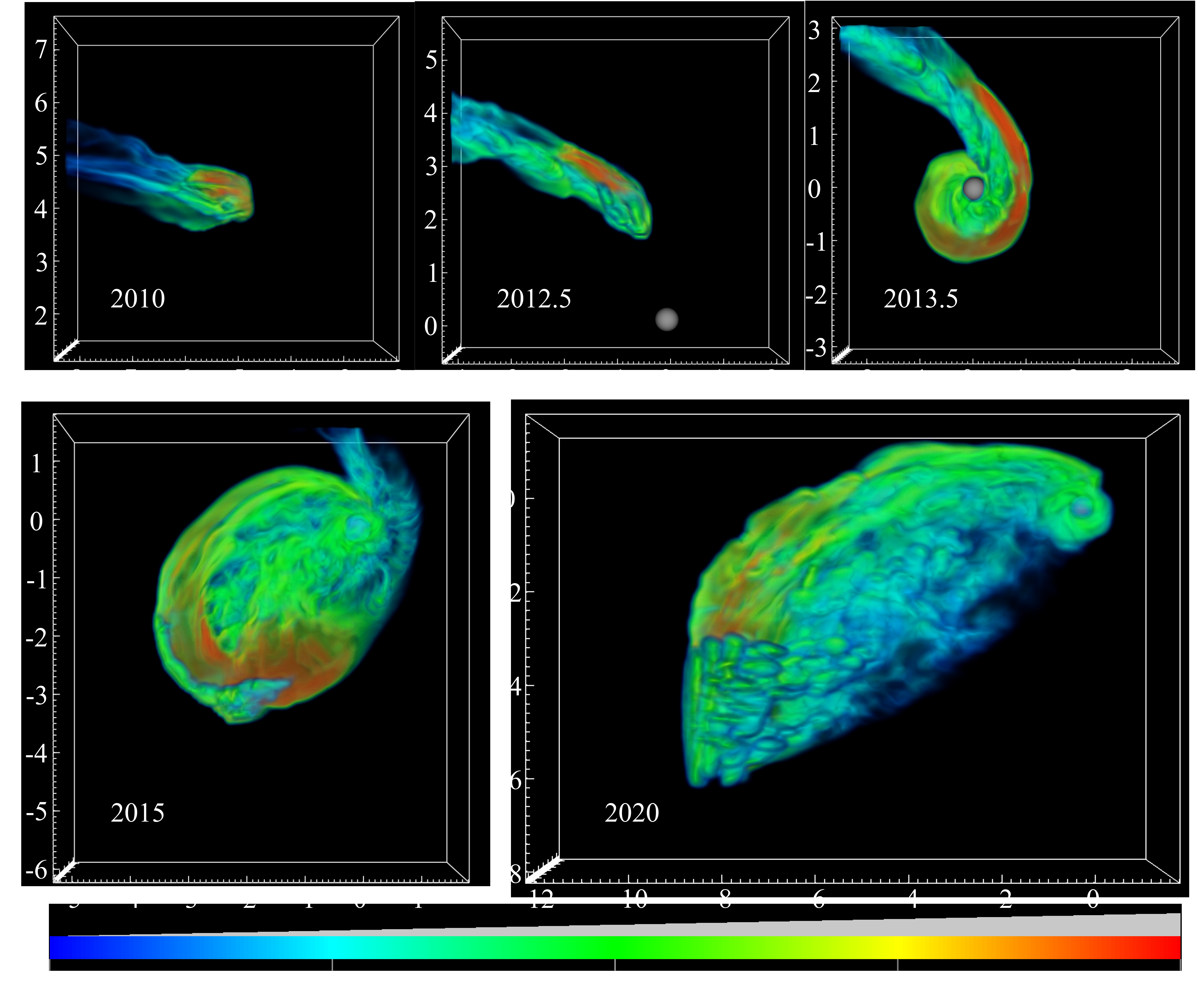}
\caption{Three-dimensional, volume visualization of cloud tracer material ${\cal T}$ for model cc\_i1\_b1\_95, spanning the period 2010 to 2020.  The small gray sphere appearing in each panel beginning in 2012.5 indicates the accretion volume, $r < r_\mathrm{acc} = 750 R_S$.  The color represents the amount of tracer (value of ${\cal T}$) in a particular region, with the transparency also dependent upon the value of ${\cal T}$ (as indicated by the slanted gray bar above the color bar).  The cloud material initially has a uniform value of ${\cal T}_0 = 2.36$.  An animated movie of this model is available with the online material.
\label{fig:tracer}}
\end{figure}

Figure \ref{fig:tracer} shows snapshots of the distribution of cloud tracer material for model cc\_i1\_b1\_95 over 10 years of its orbit, starting in 2010.  We do not show earlier epochs, since the cloud is largely spherical prior to about 2010.  As we are presenting three-dimensional simulations, we generally use volume visualizations, which allows one to see the entire simulation domain.  There is, of course, an opacity factor, so one can only see through low tracer regions.  Only in Figures \ref{fig:temperature} and \ref{fig:kappa} do we show two-dimensional slices through our data.

As expected from our estimates in Section \ref{sec:timescales}, the early phase of evolution is dominated by tidal stretching.  By 2012.5, the cloud covers an arc of roughly $1.0 \times 10^{16}$ cm ($5.3 R_c^*$), compared to widths and depths of roughly $2.0 \times 10^{15}$ cm ($1.1 R_c^*$) and $1.4 \times 10^{15}$ cm ($0.73 R_c^*$), respectively.  Ram pressure forces become important at later phases when they compress the head of the cloud and slow its orbital motion.  Hydrodynamic instabilities acting throughout the cloud history produce streams of material that drift away from the main cloud body, particularly along its inner edge.  As this material loses angular momentum relative to its Keplerian value (through interaction with the background gas), it accelerates toward the black hole ahead of the main cloud body.  By 2013.5, this material begins to penetrate within the accretion radius.  

Beginning around 2013.5, the destruction of the cloud greatly accelerates, so that, by 2015, it has separated into numerous filaments, with additional cloud material dispersed over a volume many times larger than the initial volume of the cloud.  By 2020 it is clear that very little, if any, cloud material remains on the original Keplerian trajectory.  Instead, the bulk of the cloud material has been stretched out into an extended filament, anchored at the accretion radius $r_\mathrm{acc}$.  

The overall morphology of model cc\_i1\_b1\_95 shows some similarities to the comparable two-dimensional model, CC01, presented in \citet{schartmann12} (cf. their Figure 4 vs. our Figure \ref{fig:tracer}).  Certainly the position of the bulk of the cloud material is similar in each corresponding epoch.  However, many differences are also apparent, especially in the fine-level detail of the accretion filaments.  For these, the two-dimensional simulations appear to show a very thin, delicate stream connecting the bulk of the cloud to the accretion radius, whereas the three-dimensional simulations show a much more dispersed distribution.  There are two factors contributing to this difference.  One is simply a projection effect.  If we were to take a single slice through one of our three dimensional simulation volumes, as we do in Figure \ref{fig:temperature}, we would see a much greater level of fine detail, which is lost when looking through the entire volume with opacity effects.  The other, more crucial, factor is that two dimensional simulations severely restrict the avenues by which the cloud can break up.  The extra degree of freedom in three dimensional simulations has a significant effect on the evolution of the cloud, a point we return to in Section \ref{sec:accretion}.

\subsection{Equation of State Comparison}
\label{sec:eos}

\begin{figure}
\plotone{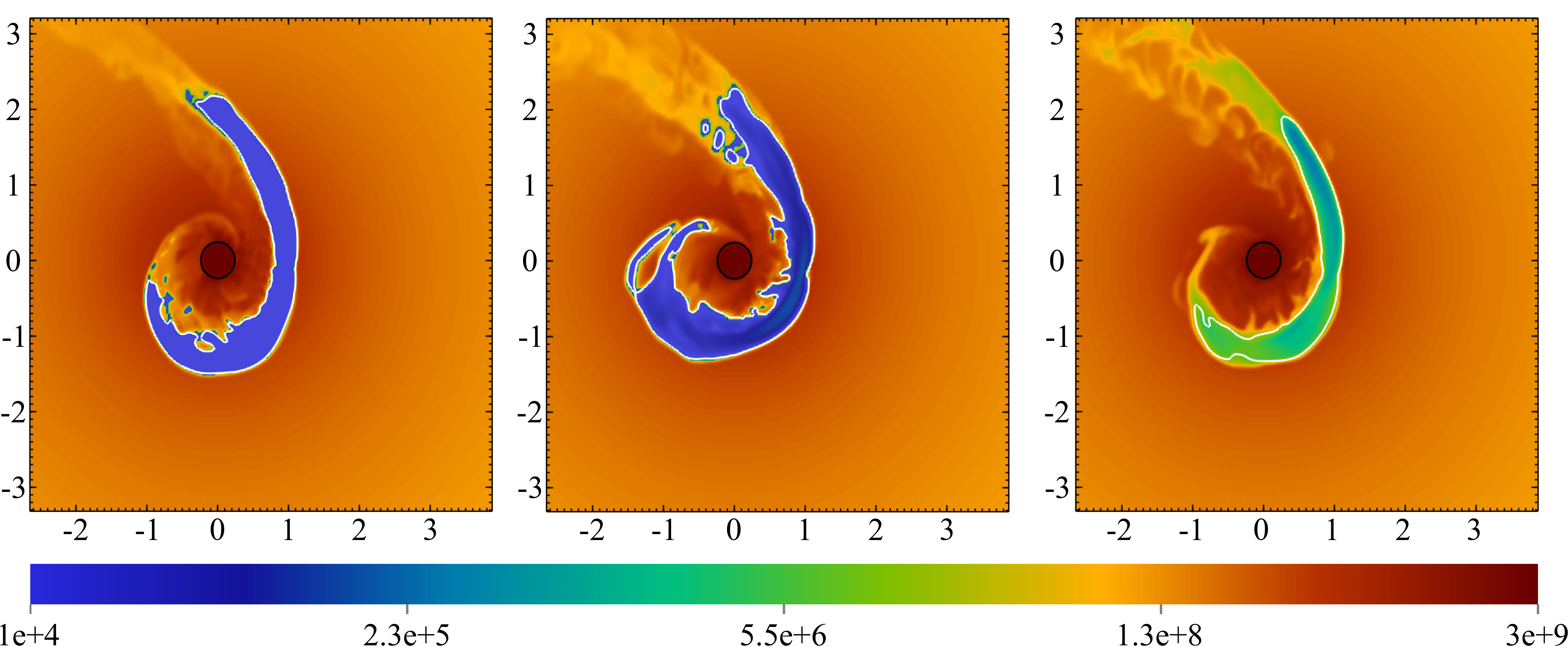}
\caption{Pseudocolor plots of temperatures for models cc\_i1\_b1\_95, cc\_i53\_b1\_95, and cc\_p53\_b1\_95 on date 2013.5.  These plots are two-dimensional slices in the $z=0$ plane.  The black circle indicates the accretion radius, $r_\mathrm{acc} = 750 R_S$, while the white contour indicates a cloud tracer level of ${\cal T} = 0.5$.  
\label{fig:temperature}}
\end{figure}

\begin{figure}
\plotone{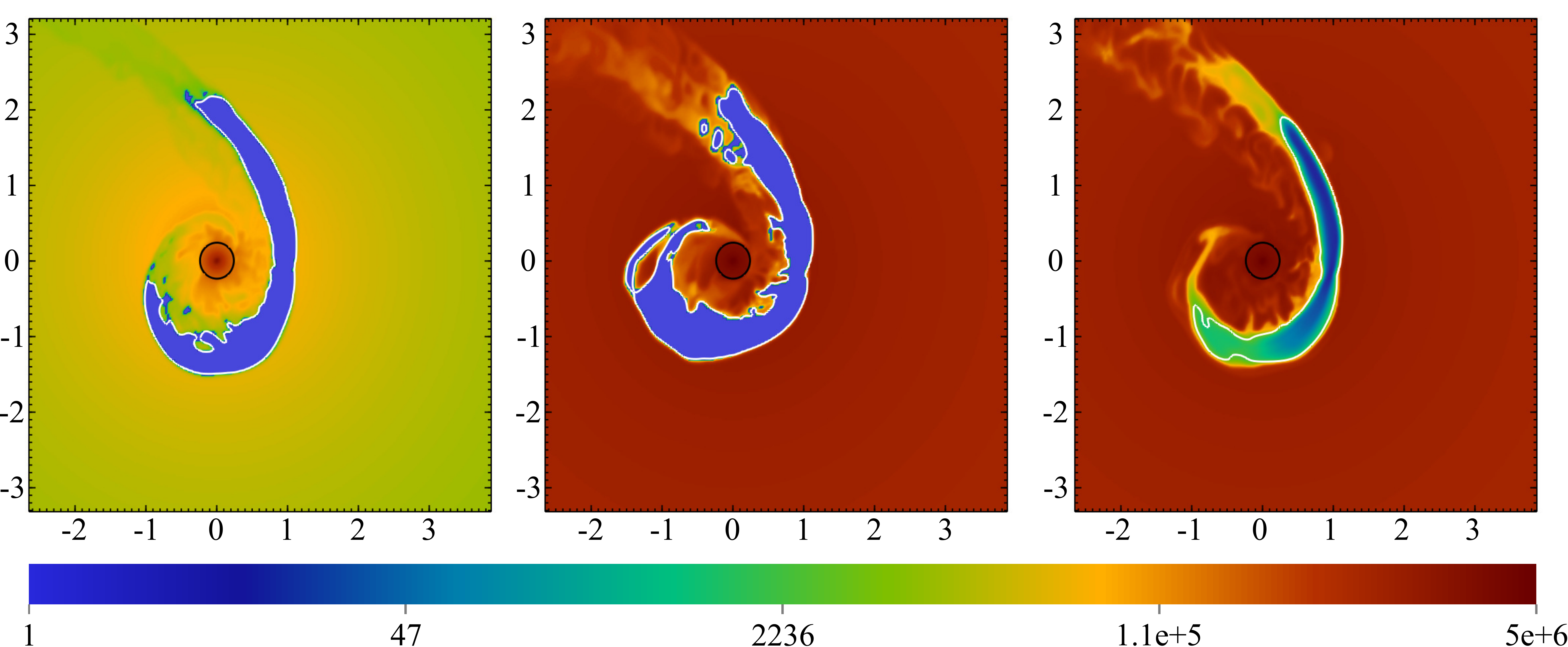}
\caption{Pseudocolor plots of the isentropic constant $\kappa = P/\rho^\Gamma$, normalized to its initial value, for models cc\_i1\_b1\_95, cc\_i53\_b1\_95, and cc\_p53\_b1\_95 on date 2013.5.  These plots are two-dimensional slices in the $z=0$ plane.  The black circle and white contours have the same meaning as in Fig. \ref{fig:temperature}.   
\label{fig:kappa}}
\end{figure}

Figure \ref{fig:temperature} shows the temperature of the cloud for our three different EOS models on date 2013.5.  The isothermal ($\Gamma = 1$) cloud model, cc\_i1\_b1\_95, of course maintains its initial temperature $T = 10^4$ K.  The isentropic ($\Gamma = 5/3$) model, cc\_i53\_b1\_95, shows some temperature variation, but the bulk of the cloud material remains below $10^5$ K.  The polytropic ($\Gamma=5/3$) model, cc\_p53\_b1\_95, on the other hand, shows considerable heating, with much of the cloud approaching $10^7$ K.  The differences between the two $\Gamma = 5/3$ simulations indicate that non-adiabatic processes (i.e. shocks) play a significant role.  This can be seen more directly by noting in Figure \ref{fig:kappa} that the isentropic constant $\kappa$ retains its initial value in models cc\_i1\_b1\_95 and cc\_i53\_b1\_95, but changes quite substantially for the more general polytropic model cc\_p53\_b1\_95.

There are also some interesting differences in the morphology of the clouds in each of these simulations. For example, Figure \ref{fig:temperature} shows a progression towards greater cloud disruption by the Kelvin-Helmoltz instability as internal heating increases. This is attributed to the smaller density differential between cloud and background gas as the cloud volume increases and its density decreases with internal heating. A smaller density differential decreases the timescale for Kelvin-Helmholtz, as shown in equation (\ref{eqn:kh}). However, the total mass lost from the cloud by this mechanism by 2013.5 is less than 3\% for all three EOS models (based on measuring the amount of cloud material no longer contained within the contours shown in Figures \ref{fig:temperature} and \ref{fig:kappa}).

The most significant difference exhibited by the three equation of state models is the disruption of the head of the cloud as it passes pericenter and begins to decelerate. Ram pressure becomes increasingly important at this time as the head of the cloud slows and the tail piles up behind. Rayleigh-Taylor instabilities are also triggered at the head of the cloud at this time as the background gas effectively accelerates into the denser cloud material. The timescales for Rayleigh-Taylor are shortest for greater acceleration or pressure differential between cloud and background gas. Hence, as shown in Figure \ref{fig:temperature}, the polytropic model is least affected by Rayleigh-Taylor disruption due to its ability to achieve pressure equilibrium with the background gas.

The polytropic equation of state model indicates that internal dynamic heating of the cloud is considerable.  Nevertheless, the extreme efficiency of radiative cooling is more than enough to keep the temperature of the actual cloud close to $10^4$ K.  The isothermal model should, therefore, represent the most realistic simulation of the cloud's evolution.

\subsection{Background Comparison}
\label{sec:background}

\begin{figure}
\plotone{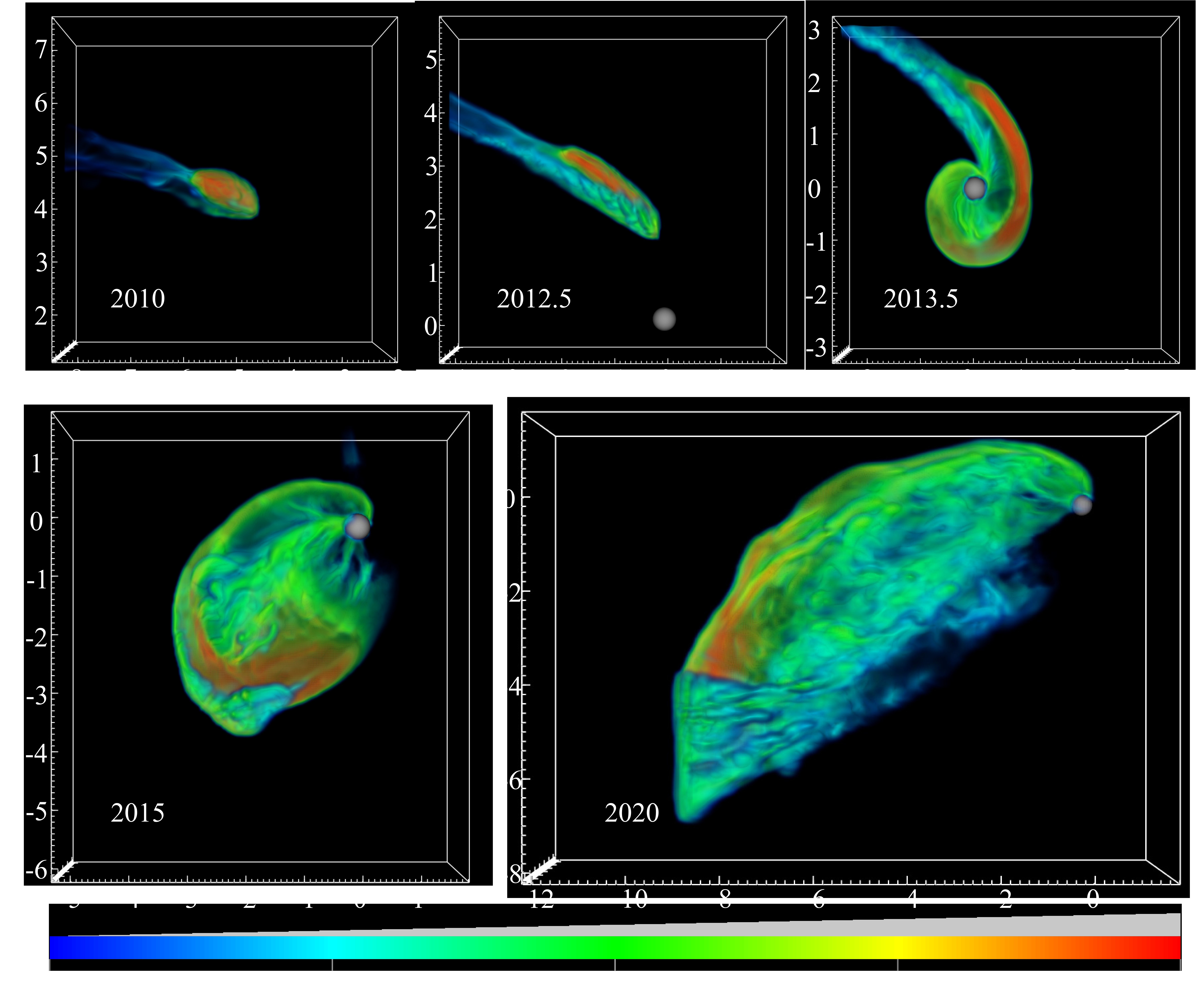}
\caption{Same as Fig. \ref{fig:tracer}, but for the alternate background model, cc\_i1\_b2\_95.
\label{fig:b2}}
\end{figure}

The convectively stable background profile given by equations (\ref{eq:rho_back_2}) - (\ref{eq:T_back_2}) does not appear to lead to a dramatically different cloud morphology at early (2012.5 and earlier) nor late (2015 and later) times, as seen by comparing model cc\_i1\_b2\_95 in Figure \ref{fig:b2} with model cc\_i1\_b1\_95 in Figure \ref{fig:tracer}; there is, however, some difference in the distribution of cloud material around the accretion sphere in 2013.5.  Nevertheless, we find that the average mass accretion rate in cc\_i1\_b2\_95 is only about 10\% lower than cc\_i1\_b1\_95, as shown in Table \ref{tab:models} and Figure \ref{fig:accretionrate}.  This is perhaps not surprising since we artificially reset both backgrounds to their initial profiles after each hydro cycle.  This artificially stabilizes the convectively unstable background.

Still, the small differences between the two background models can give us some estimate of the level of sensitivity of our results to realistic variations in the true galactic center environment.  Specifically, our work probes the sensitivity to density and temperature.  However, other properties of the galactic center may also be important.  For example, radiation and magnetic fields might affect the evolution of G2, and neither is considered in this work.

\subsection{Start Date - 1944.6 vs. 1995.5}
\label{sec:start}

\begin{figure}
\plotone{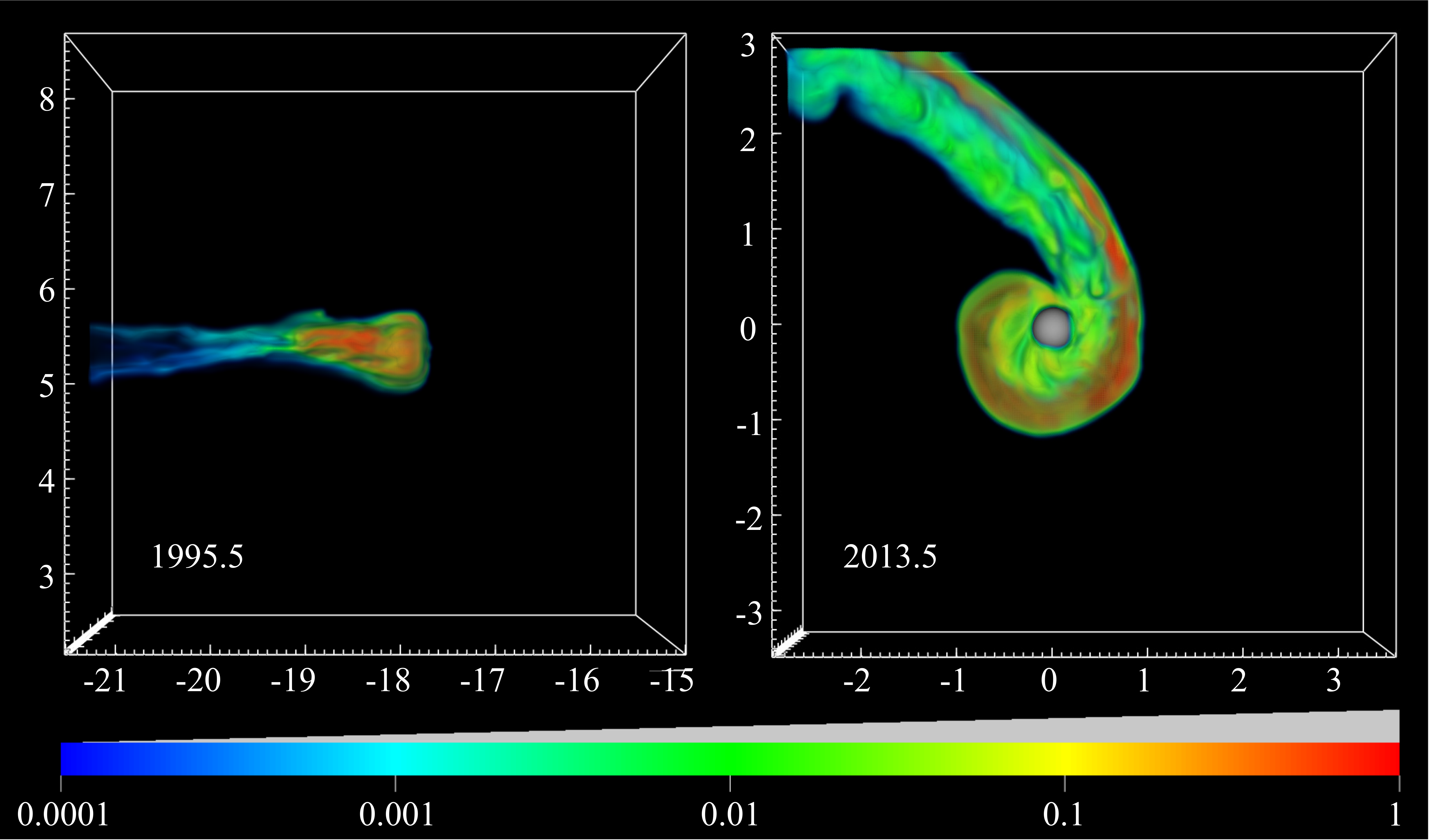}
\caption{Same as Fig. \ref{fig:tracer}, but for the model that started in 1944.6, cc\_i1\_b1\_44, shown on dates 1995.5 and 2013.5.
\label{fig:1944}}
\end{figure}

All but one of our simulations began with a start date of 1995.5, shortly before the discovery of G2.  For the other, we started the cloud from what would be the apocenter of its present orbit.  The cloud in this simulation, cc\_i1\_b1\_44, obviously has much longer to evolve and traverses a much greater distance through the galactic center.  In Figure \ref{fig:1944}, we show what the cloud from this model would have looked like on 1995.5.  It is clear that tidal forces are already beginning to stretch the cloud and ram pressure forces are blunting its leading edge, as expected based on our calculations in Section \ref{sec:timescales}.  Clearly if G2 started as a spherical cloud at apocenter, then its shape should have changed significantly by the time of its discovery.  Ram pressure and hydrodynamic instabilities are also stripping some material from the edges of the cloud, generating a tail of tracer material. The mass loss to this point, however, is negligible; the cloud still retains 99.95\% of its material.  Nevertheless, the earlier seeding of these instabilities ultimately leads to a greater disruption of the cloud, starting at pericenter passage (cf. Figure \ref{fig:1944}).  

Our results support the conclusions of \citet{burkert12} and \citet{schartmann12} that G2 must have either formed shortly before its discovery or be part of a larger, more extended structure.  Otherwise, it should have had a much more elongated appearance at the time of its 2011 observation.  It is also clear that G2 will not survive even a single passage through the galactic center; the environment is just too hostile for a cloud of its size.

\subsection{Mass Accretion}
\label{sec:accretion}

\begin{figure}
\plotone{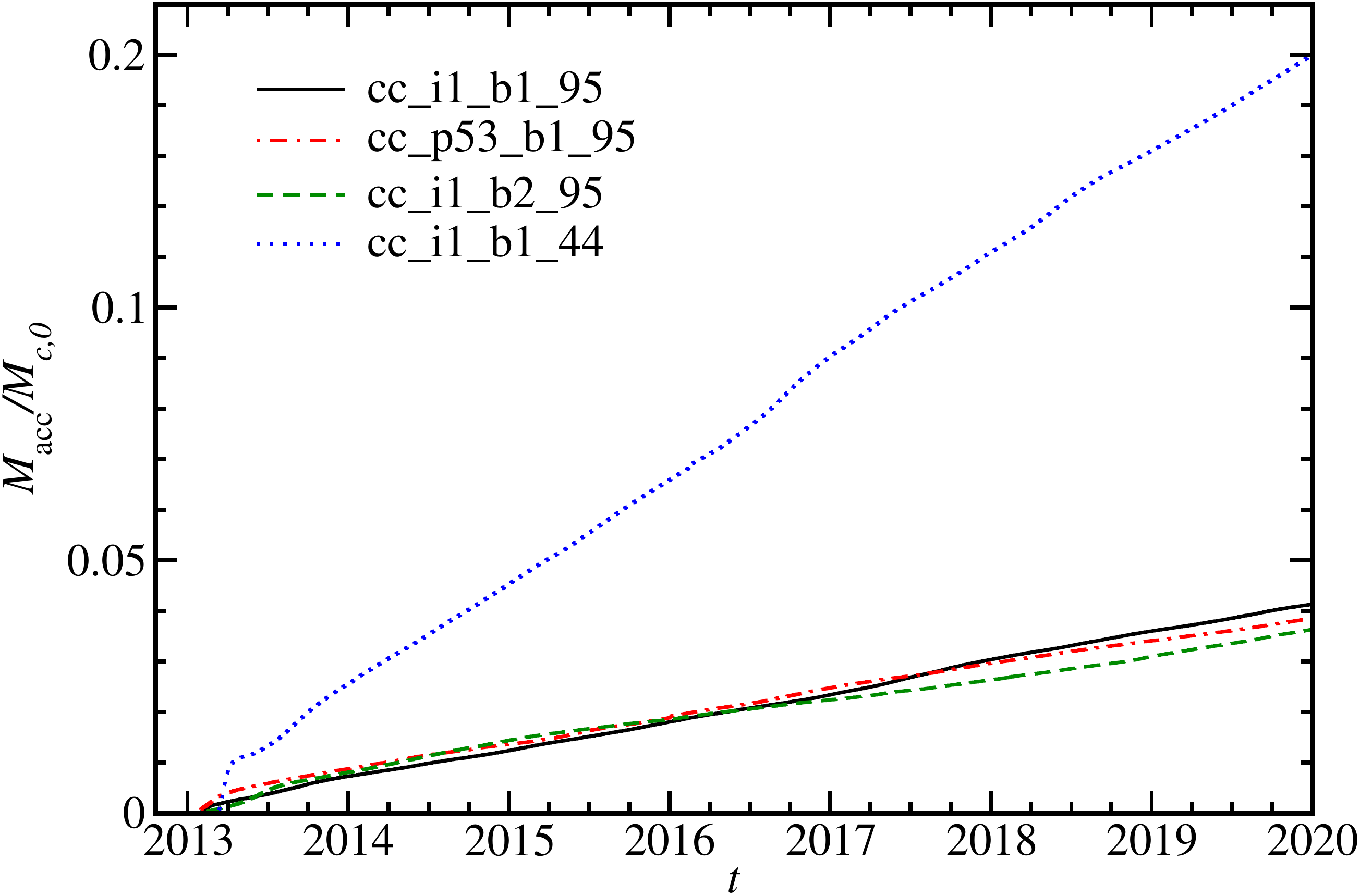}
\caption{Plot of the total amount of cloud material accreted through $r_\mathrm{acc}$ as a function of time, represented as a fraction of the initial cloud mass, for four models.  
\label{fig:accretedmass}}
\end{figure}

During the simulations, we trace the mass inflow through the accretion radius, $r_\mathrm{acc} = 750 R_S$.  Only the accreted mass that was originally part of the cloud is recorded.  The resulting total accreted mass is plotted against time in Figure \ref{fig:accretedmass} as a fraction of the initial cloud mass for four of the models (we leave off the isentropic model to prevent overcrowding).  Cloud material first begins to accrete in 2013.  By 2020, roughly 4\% of the cloud material has accreted in most simulations, giving a baseline mass accretion rate of $5 \times 10^{-8} M_\odot~\mathrm{yr}^{-1}$.  

\begin{figure}
\plotone{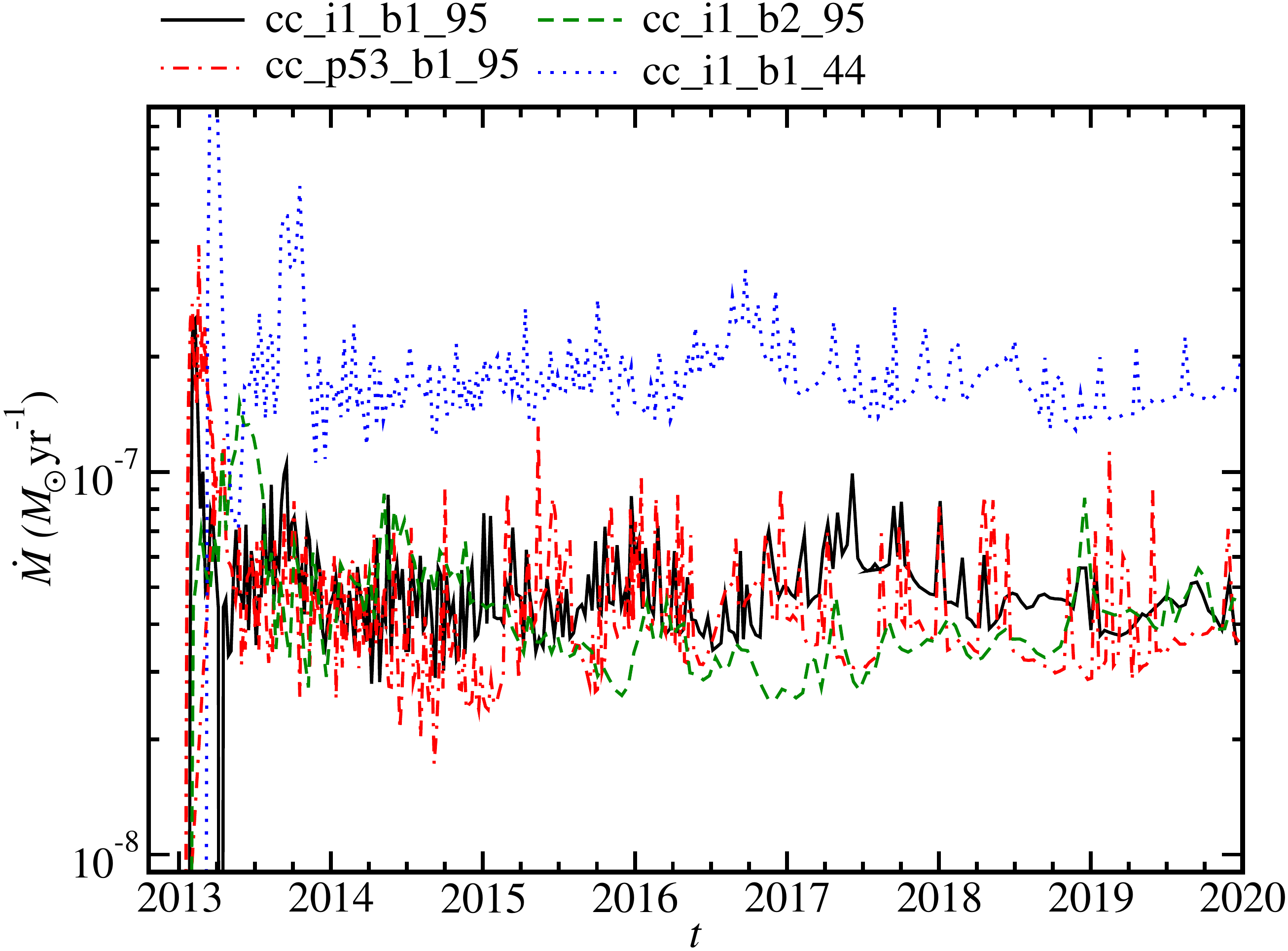}
\caption{Plot of the instantaneous mass accretion rate through $r_\mathrm{acc}$ for four models.  The data are sampled every 100th cycle of the hydrodynamics solver, giving an effective time sampling of $\approx 0.03~\mathrm{yr} = 10~\mathrm{d}$.
\label{fig:accretionrate}}
\end{figure}

Time-resolved mass accretion rates are shown in Figure \ref{fig:accretionrate}.  Averages and standard deviations are also reported in Table \ref{tab:models}.  During the simulations, the time-averaged accretion rate exhibits only stochastic changes over the seven year period we monitor.  The mean accretion rate in all models remains fairly constant, and for model cc\_i1\_b1\_95 is comparable to the analogous two-dimensional model, CC01, from \citet{schartmann12}.  However, whereas the two-dimensional simulations show variability up to an order of magnitude on timescales as short as $\sim 1$ month, we observe variations no larger than about a factor of three on similar timescales.  This is another example of where the detailed evolution differs significantly in two versus three dimensions.

One goal of this paper is to estimate how this mass accretion will affect the future activity of Sgr A*.  To do this, we need to compare our measured accretion rates against the current estimated mass accretion rate in the galactic center.  Studies \citep[e.g.][]{blandford99,yuan03} suggest the following radial dependence
\begin{equation}
\dot{M}(r) = \dot{M}_\mathrm{Bondi} \left( \frac{r}{r_\mathrm{Bondi}} \right)^{0.27}~,
\end{equation}
with $\dot{M}_\mathrm{Bondi} = 10^{-5} M_\odot~\mathrm{yr}^{-1}$ \citep{yuan03} at $r_\mathrm{Bondi}$.  This implies that the current mass accretion rate at $r = r_\mathrm{acc}$, where we make our measurements, should be $3.8 \times 10^{-6} M_\odot~\mathrm{yr}^{-1}$.  In this case, the capture of G2 material onto Sgr A* would only boost the observed mass accretion rate by about 1-5\%, depending on which of our models we consider.  

The \citet{yuan03} estimate, though, assumes that the gas accretes as a radiatively inefficient accretion flow (RIAF).  It could be that the cloud material, being cooler, denser, and having a coherent angular momentum, may accrete much more efficiently onto the black hole.  If that is the case, then our observed mass accretion rate, $5-19 \times 10^{-8} M_\odot~\mathrm{yr}^{-1}$, may lead to a significant increase in the mass accretion rate onto Sgr A*, currently estimated to be in the range $2\times10^{-9}<\dot{M} <2\times10^{-7} M_\odot~\mathrm{yr}^{-1}$ \citep{aitken00, bower03, marrone07}.  Since modeling the galactic center mass accretion rate relies on a number of uncertain assumptions about energy transport and radiative efficiency, it will be extremely useful to monitor the break up of G2 to better understand how Sgr A* is fed.

\subsection{Luminosity}
\label{sec:luminosity}

\begin{figure}
\plotone{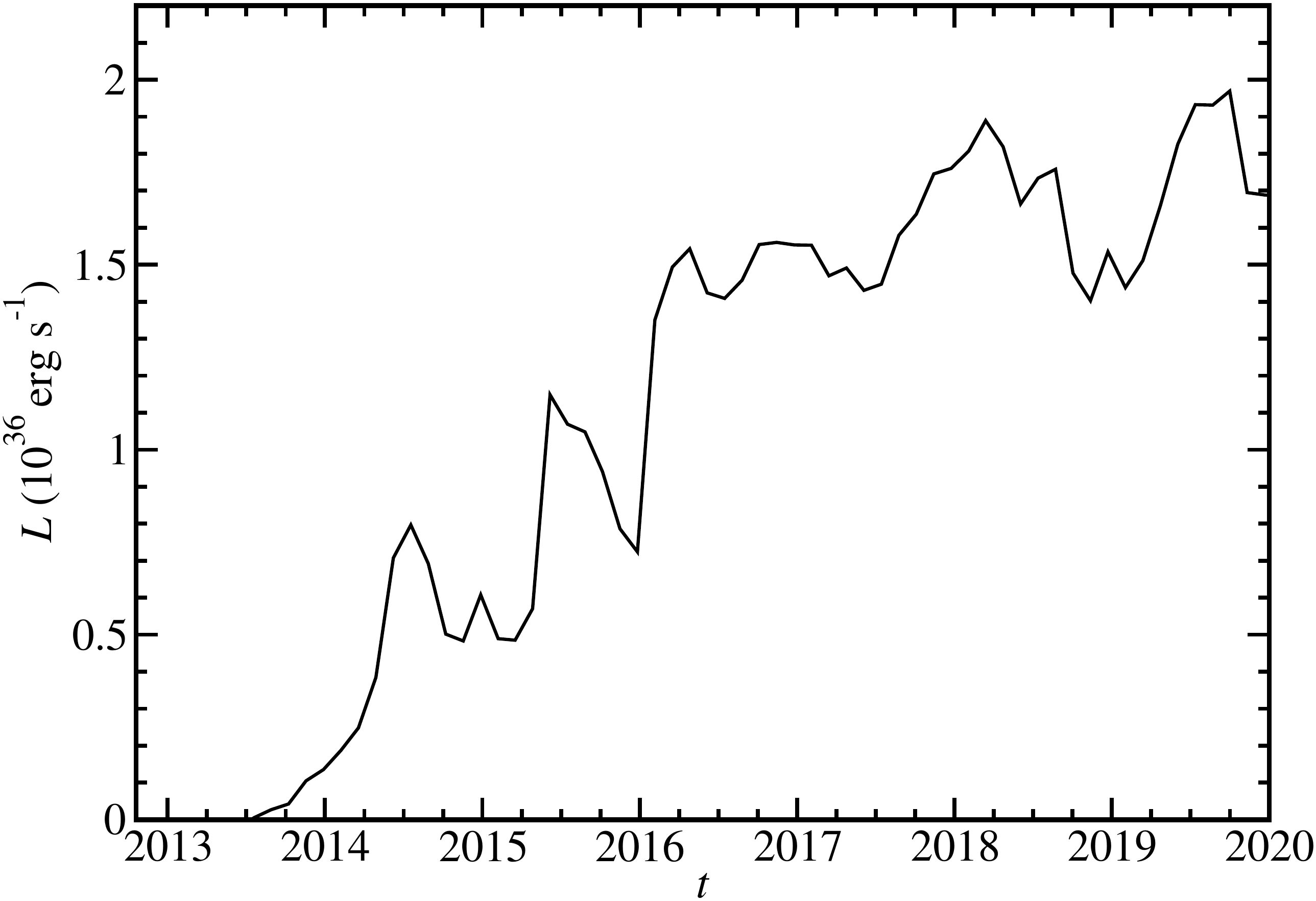}
\caption{Light curve for model cc\_i1\_b1\_95, estimated from the dissipation of the gravitational potential plus kinetic energy of the cloud.
\label{fig:luminosity}}
\end{figure}

We can get a crude estimate (upper limit) of the luminosity associated with the disruption of G2 by tracking the amount of cloud energy (gravitational potential plus kinetic) that is dissipated during our simulations.  Assuming that all of this energy is radiated away instantaneously, we can track its luminosity as a function of time as shown in Figure \ref{fig:luminosity} for model cc\_i1\_b1\_95.  The resulting value, $\sim 10^{36}~\mathrm{erg~s}^{-1}$, is roughly consistent with the predicted pericenter luminosity from \citet{gillessen12}.  

\section{Conclusions}
\label{sec:conclusions}

The cloud G2 is currently approaching Sgr A* and in 2013.5, will pass it at a pericenter distance of 3100 times the Schwarzschild radius, corresponding to $4.0\times10^{15}$ cm. This provides a unique opportunity to investigate directly the disruption of a cold gas clump by its gravitational interaction with the black hole and hydrodynamic interaction with the galactic center environment.  Clear evidence for tidal velocity shearing and stretching has already been detected \citep{burkert12}.

In this paper we presented five different three-dimensional, numerical, moving mesh simulations illustrating possible future evolutionary tracks for G2.  Our results generally support previous conclusions \citep{burkert12,schartmann12} that G2 must have been formed recently or must be only a small piece of a larger cloud complex.  In any case, a cloud like the ones we simulated would not be expected to survive long (apparently not even a single pericenter passage) within the galactic center.  

In all of our simulations we find a roughly similar morphology for the cloud: Tidal stretching and Kelvin-Helmholtz effects become evident first; then, around the time of pericenter passage, ram pressure and Rayleigh-Taylor instabilities additionally help to disrupt the cloud, which begins to fragment; as the cloud disperses over a larger volume, it begins to lose significant amounts of angular momentum through its interactions with the background gas; the cloud stretches into a stream of gas feeding the black hole (for us the accretion volume $r<r_\mathrm{acc}$) at $5-19 \times 10^{-8} M_\odot~\mathrm{yr}^{-1}$.  This is consistent with the values seen in the comparable two-dimensional simulation CC01 of \citet{schartmann12}, although the variability of that simulation was up to a factor of three larger.  Looking in more detail, we note some important differences among our models.  The non-isothermal models, particularly the polytropic one, showed significantly higher cloud temperatures than the more realistic isothermal ones (which accounted for radiative cooling of the cloud).  Furthermore, the isentropic model yielded a significantly higher mass accretion rate than the isothermal (and even polytropic) models. The model with the denser (convectively stable) background gave a somewhat smaller mass accretion rate than the default background.  Finally, not surprisingly, model cc\_i1\_b1\_44 illustrated that starting the simulation earlier leads to greater disruption of the cloud by the time of pericenter passage.

It is uncertain how much the disruption of G2 will affect the mass accretion feeding Sgr A*.  The accretion rate observed in our simulations is comparable to the currently estimated rate in the immediate vicinity of Sgr A*.  But at the radius where we are actually measuring the accretion, it only represents an increase of 1-5\% over the currently estimated feeding rate.  Such a small change in $\dot{M}$ may not produce any measurable changes in the Sgr A* emission (although see \citet{moscibrodzka12} for examples of what larger accretion rate changes might do to the X-ray, infrared, and millimeter emission from Sgr A*).  Ultimately it will depend on how efficiently the G2 material can get from the radius where our simulations cut off down to Sgr A*.  Regardless, the break up of G2 will provide an unprecedented opportunity to study accretion physics in the galactic center.  Furthermore, our crude luminosity estimate of $\sim 10^{36}~\mathrm{erg~s}^{-1}$ suggests this event should be observable over the next several years.

Concerning possible future improvements to our work, we have not explored the following alternate scenarios: the ``spherical shell'' of \citet{burkert12} and \citet{schartmann12}, the ``hidden-star/proto-planetary disk'' of \citet{murray-clay11}, nor the ``photoevaporating, disturbed stellar disk'' of \citet{miralda12}.  In the first, G2 is just the tip of a much larger, more massive shell of gas.  This scenario may be required to explain the lower surface brightness, cone-like structure that accompanies G2 \citep{gillessen12}.  The second scenario proposes that G2 is a dense, proto-planetary disk bound to a low-mass star, which was scattered from the disk of young stars orbiting Sgr A*.  The third suggests instead that G2 is a cloud of photoevaporated gas originating from a disk around a star that was previously disrupted by a stellar mass black hole and now produces a cloud at every periastron passage.  We have also not considered thermal conduction.  This is likely the next most important physical influence beyond those considered in this work.  Thermal conduction will be especially important at late times as the cloud breaks up.  Magnetic fields may also play an important role.  For example, magnetic fields can affect the growth and development of hydrodynamic instabilities along the cloud surface \citep[e.g.][]{fragile05}.

\begin{acknowledgments}
We want to thank the anonymous referee for useful comments on this paper.  This work was supported in part by the National Science Foundation under Grant No. NSF PHY11-25915 and by NSF Cooperative Agreement Number EPS-0919440 that included computing time on the Clemson University Palmetto Cluster.  The work by PA and SDM was performed under the auspices of the U.S. Department of Energy by Lawrence Livermore National Laboratory under Contract DE-AC52-AC52-07NA27344.
\end{acknowledgments}

\appendix

\section{Moving Mesh}
\label{sec:mesh}

As this is the first introduction of moving meshes implemented in the {\em Cosmos++} code, we present a brief description of the covariant framework from which the equations are derived. A more formal presentation of numerical methods and code tests will be considered in another paper. Throughout this Appendix we use standard index notation in which repeated indices represent summations over spatial components, and the raising and lowering of indices (to contravariant and covariant components) is done with the spatial metric, e.g. $v_\alpha=g_{\alpha\beta}v^\beta$. We also adopt the convention in non-relativistic work of using Latin indices ($i,j,k$) to represent quantities measured in Cartesian coordinates, and Greek indices ($\alpha,\beta,\gamma$) for generalized curvilinear coordinates. The indices run over the three spatial dimensions (they do not include the time component).  

Before presenting the equations, we draw an analogy with general relativistic magnetohydrodynamics (MHD) to define state variables and write the stress-energy tensor for a perfect fluid (and ideal MHD) as a linear combination of the hydrodynamic $T^{\alpha\beta}_{H}$ and magnetic contributions $T^{\alpha\beta}_{B}$:
\begin{equation}
T^{\alpha\beta} = T^{\alpha\beta}_{H} + T^{\alpha\beta}_{B} 
       = \rho v^\alpha v^\beta + (P+P_B) g^{\alpha\beta} 
         + Q^{\alpha\beta} - b^\alpha b^\beta ~,
\label{eqn:tmn}
\end{equation}
where $\rho$ is the fluid mass density, $v^\alpha$ is the contravariant fluid velocity, $P$ is the fluid pressure (for an ideal gas $P=(\Gamma-1)e$ where $e$ is the fluid internal energy density and $\Gamma$ is the adiabatic index), $b^\alpha$ is the magnetic field, $P_B = g_{\alpha\beta}b^\alpha b^\beta/2$ is the magnetic pressure, $Q^{\alpha\beta}$ is the tensor artificial viscosity used for capturing shocks, and $g_{\alpha\beta}$ is the spatial 3-metric tensor for general curvilinear coordinates. Although we do not model magnetic fields in this paper, we include the full MHD equations for completeness and future reference.

We begin by writing in flux-conserving form the evolution equations for mass, internal energy, momentum, and magnetic induction in Cartesian coordinates ($x^i$) as:
\begin{equation}
 \frac{\partial\rho}{\partial t} +
 \frac{\partial (\rho v^i)}{\partial x^i} = 0 ~,  
      \label{eqn:av_de}
\end{equation}
\begin{equation}
 \frac{\partial e}{\partial t} +
 \frac{\partial (ev^i)}{\partial x^i} =
      -\left(P \delta^j_i + Q^j_i\right) \frac{\partial v^i}{\partial x^j} ~,
      \label{eqn:av_en}
\end{equation}
\begin{equation}
 \frac{\partial s_j}{\partial t} + 
 \frac{\partial (s_j v^i)}{\partial x^i} =
      -\frac{\partial}{\partial x^i} \left[(P+P_b)~\delta^i_j + Q^i_j - b^i b_j\right]
      -\rho\frac{\partial\phi}{\partial x^j} ~,
      \label{eqn:av_mom}
\end{equation}
\begin{equation}
 \frac{\partial b^j}{\partial t} +
 \frac{\partial (b^j v^i)}{\partial x^i} =
      b^i \frac{\partial v^j}{\partial x^i} 
      +\delta^{ij} \frac{\partial \psi}{\partial x^i} ~,
      \label{eqn:av_ind}
\end{equation}
where $\delta^i_j$ is the Kronecker delta tensor, and $\delta_{ij}=g_{ij}$ is effectively the flat metric in Cartesian coordinates. Additionally $\psi$ is a scalar potential introduced as a divergence cleanser to maintain a divergence-free magnetic field ($\partial_i b^i = 0$), and $\phi$ is the gravitational potential satisfying Poisson's equation
\begin{equation}
  \frac{\partial^2\phi}{(\partial x^i)^2} = 4\pi G\rho ~.
  \label{eqn:poisson}
\end{equation}

Next we consider an arbitrary time-dependent transformation into generalized coordinates $\xi^\alpha$, with $\boldsymbol{x}(\boldsymbol{\xi},t)$, triad $e^i_\alpha = \partial x^i/\partial\xi^\alpha$, spatial 3-metric $g_{\alpha\beta}=e^i_\alpha e^j_\beta \delta_{ij}$, metric determinant $\sqrt{g}\equiv \mathrm{Det}(||g_{\alpha\beta}||)$, and grid velocity $\dot{x}^i = \partial x^i/\partial t|_\xi = V_g^i$. Applying standard transformation rules for gradients and vector quantities in general, the MHD equations take the following form in moving generalized coordinates:
\begin{equation}
 \frac{\partial \sqrt{g} \rho}{\partial t} +
 \frac{\partial}{\partial \xi^\alpha} 
      \left[\sqrt{g} \rho e^\alpha_i (v^i - V_g^i)\right] = 0 ~,  
 \label{eqn:av_de_covC}
\end{equation}
\begin{equation}
 \frac{\partial \sqrt{g} e}{\partial t} +
 \frac{\partial}{\partial \xi^\alpha} 
      \left[\sqrt{g} e e^\alpha_i  (v^i - V_g^i)\right] =
      -\left(P \delta^\alpha_\beta + Q^\alpha_\beta\right) 
       \frac{\partial(\sqrt{g}~v^\beta)}{\partial \xi^\alpha} ~,
      \label{eqn:av_en_covC}
\end{equation}
\begin{eqnarray}
 \frac{\partial \sqrt{g} s_j}{\partial t} + 
 \frac{\partial}{\partial \xi^\alpha} 
      \left[\sqrt{g} s_j e^\alpha_i  (v^i - V_g^i)\right] &=&
     -\frac{\partial}{\partial \xi^\alpha}  
      \left\{\sqrt{g} e^\beta_j \left[ (P+P_b)\delta^\alpha_\beta
                      +Q^\alpha_\beta - b^\alpha b_\beta\right]\right\} \nonumber\\
    &-& \sqrt{g} \rho e^\beta_j \frac{\partial\phi}{\partial x^\beta} ~,
    \label{eqn:av_mom_covC}
\end{eqnarray}
\begin{equation}
 \frac{\partial \sqrt{g} b^j}{\partial t} +
 \frac{\partial}{\partial \xi^\alpha} 
       \left[\sqrt{g} b^j e^\alpha_i  (v^i - V_g^i)\right]  =
       \sqrt{g} b^\alpha \frac{\partial v^j}{\partial \xi^\alpha} 
      +\sqrt{g} g^{\alpha\beta} e^j_\beta \frac{\partial \psi}{\partial \xi^\alpha} ~.
      \label{eqn:av_ind_covC}
\end{equation}
\begin{equation}
  \frac{\partial}{\partial \xi^\alpha}
       \left(\sqrt{g} g^{\alpha\beta}
       \frac{\partial\phi}{\partial \xi^\beta}\right) = 4\pi G\sqrt{g}\rho ~,
  \label{eqn:poisson_cov}
\end{equation}
In this form the evolved vector fields (velocity, momentum and magnetic field) are defined in Cartesian coordinates ($x^i$).

Taking these transformations a step further, we assume the triad matrix is a product of two transformation bases
\begin{equation}
   e^j_\alpha = e^j_{\widehat\beta}~e^{\widehat\beta}_\alpha ~,
\end{equation}
where $e^j_{\widehat\beta}$ is a static (time independent) transformation between Cartesian coordinates and any other fixed frame coordinate set ($\xi^{\widehat\alpha}$), for example spherical or cylindrical coordinates. The second triad $e^{\widehat\beta}_\alpha$ is a dynamical (time dependent) transformation between the static fixed frame coordinate system (spherical, cylindrical) and generalized curvilinear coordinates. This decomposition allows for grid motion in more general topological ``background'' grids besides Cartesian. It affects only the vector conservation equations, allowing vector quantities to be expressed in the most convenient (symmetry respecting) triad. Hence only the momentum and magnetic field equations (\ref{eqn:av_mom_covC}) and (\ref{eqn:av_ind_covC}) are modified under this transformation as
\begin{eqnarray}
 \frac{\partial \sqrt{g} s_{\widehat\beta}}{\partial t} + 
 \frac{\partial}{\partial \xi^\alpha} 
      \left[\sqrt{g} s_{\widehat\beta} e^\alpha_{\widehat\sigma} 
                     (v^{\widehat\sigma} - V_g^{\widehat\sigma})\right] &=&
     -\frac{\partial}{\partial \xi^\alpha}  
      \left\{\sqrt{g} e^\sigma_{\widehat\beta} \left[ (P+P_b)\delta^\alpha_\sigma
                      +Q^\alpha_\sigma - b^\alpha b_\sigma\right]\right\} \nonumber\\
    &-& \sqrt{g} \rho e^\alpha_{\widehat\beta} \frac{\partial\phi}{\partial \xi^\alpha}
       +\sqrt{g} e^\alpha_{\widehat\gamma} 
                 \left( T^{\widehat\sigma\widehat\gamma} 
                       -s^{\widehat\sigma} V_g^{\widehat\gamma}\right)
                 \Gamma_{\widehat\sigma\alpha\widehat\beta} ~,
    \label{eqn:av_mom_covC2}
\end{eqnarray}
\begin{equation}
 \frac{\partial \sqrt{g} b^{\widehat\beta}}{\partial t} +
 \frac{\partial}{\partial \xi^\alpha} 
       \left[\sqrt{g} b^{\widehat\beta} e^\alpha_{\widehat\sigma} 
                      (v^{\widehat\sigma} - V_g^{\widehat\sigma})\right]  =
       \sqrt{g} b^\alpha \frac{\partial v^{\widehat\beta}}{\partial \xi^\alpha} 
      +\sqrt{g} b^{\widehat\sigma} (e^\alpha_{\widehat\gamma} V_g^{\widehat\gamma})
                \Gamma^{\widehat\beta}_{\alpha\widehat\sigma}
      +\sqrt{g} g^{\alpha\sigma} e^{\widehat\beta}_\sigma 
                \frac{\partial \psi}{\partial \xi^\alpha} ~.
      \label{eqn:av_ind_covC2}
\end{equation}
The mixed index (hatted and unhatted) term $\Gamma_{\widehat\sigma\alpha\widehat\beta}$ in equation (\ref{eqn:av_mom_covC2}) is actually $\delta_{ij} e^i_{\widehat\sigma} \partial_\alpha e^j_{\widehat\beta}$, and $ \Gamma^{\widehat\beta}_{\alpha\widehat\sigma}$ in (\ref{eqn:av_ind_covC2}) is $e^{\widehat\beta}_j \partial_\alpha e^j_{\widehat\sigma}$, which follow from the Christoffel/triad gradient identities. Both of these terms are time-independent and can be evaluated as static triad Christoffel symbols at the initial time when dynamical and static triad coordinate systems are identical.


\end{document}